\documentclass[12pt]{article}

\usepackage{amsmath,amssymb}
\usepackage{xspace}  
\usepackage{epsfig}  
  
\usepackage{graphicx}               

\newcommand{\ap}{\alpha'}

\newcommand{\nc}{\newcommand}  
  
\nc{\beq}{\begin{equation}}  
\nc{\eeq}{\end{equation}}  
\nc{\beqa}{\begin{eqnarray}}  
\nc{\eeqa}{\end{eqnarray}}  
\nc{\bea}{\begin{eqnarray}}  
\nc{\eea}{\end{eqnarray}}  
\nc{\ra}{\rightarrow}  
\nc{\lsim}{\begin{array}{c}\,\sim\vspace{-21pt}\\< \end{array}}  
\nc{\gsim}{\begin{array}{c}\sim\vspace{-21pt}\\> \end{array}}  
  
\nc{\LL}{L}  
\nc{\vv}{\tilde{v}}  
\nc{\GG}{\widetilde{G}}

\makeatletter
\@addtoreset{equation}{section}
\makeatother

\title{  
\vspace*{-2.3cm}  
\begin{flushright}  
\normalsize{  
CERN-PH-TH/2005-247\\
LPTHE-05-33\\
UAB-FT-595
  }  
\end{flushright}  
\vspace{1.5cm}  
\Large \textbf{\sc Split extended supersymmetry from\\ intersecting
branes}\vspace*{1.0cm} \author{\large\textbf{Ignatios
Antoniadis~$^{a,}$\footnote{On leave of absence from CPHT, Ecole
Polytechnique, UMR du CNRS 7644}}, \textbf{Karim Benakli~$^b$},
\textbf{Antonio Delgado~$^a$}, \\[0.3cm] \textbf{Mariano
Quir\'os~$^c$} and \textbf{Marc Tuckmantel~$^{a,d}$}\\ \\[0.5cm]
$^a$\normalsize\emph{TH-Division, CERN, 1211 Geneva, Switzerland}\\
$^b$\normalsize\emph{Laboratoire de Physique Th\'eorique et Hautes
Energies}\\ 
\normalsize\emph{Universit\'es de Paris VI et VII, France}\\
$^c$\normalsize\emph{Instituci\'o Catalana de Recerca i Estudis
Avan\c{c}ats (ICREA)}\\ \normalsize\emph{Theoretical Physics
Group, IFAE/UAB, E-08193 Bellaterra, Barcelona, Spain}\\
$^d$\normalsize\emph{Institut f\"ur Theoretische Physik, ETH
H\"ongerberg, 8093 Z\"urich, Switzerland }} \date{}}

\begin{document}  
\setcounter{page}{0}  
\maketitle  
\begin{abstract}  

We study string realizations of split extended supersymmetry, recently
proposed in hep-ph/0507192.  Supersymmetry is broken by small
($\epsilon $) deformations of intersection angles of $D$-branes giving
tree-level masses of order $m_0^2\sim \epsilon M_s^2$, where $M_s$ is
the string scale, to localized scalars. We show through an explicit
one-loop string amplitude computation that gauginos acquire
hierarchically smaller Dirac masses $m_{1/2}^D \sim m_0^2/M_s$. We
also evaluate the one-loop Higgsino mass, $\mu$, and show that, in the
absence of tree-level contributions, it behaves as $\mu\sim
m_0^4/M_s^3$. Finally we discuss an alternative suppression of scales
using large extra dimensions. The latter is illustrated, for the case
where the gauge bosons appear in $N=4$ representations, by an explicit
string model with Standard Model gauge group, three generations of
quarks and leptons and gauge coupling unification.

\end{abstract}  
  
\thispagestyle{empty}  

\newpage

\setcounter{page}{1}

\baselineskip18pt

\section{\sc Introduction}  

Implementing the idea of split supersymmetry~\cite{split} in string
theory is straightforward~\cite{Antoniadis:2004dt}. The appropriate
framework is type I theory~\cite{Angelantonj:2002ct} compactified in
four dimensions in the presence of constant internal magnetic
fields~\cite{Bachas:1995ik, Angelantonj:2000hi}, or equivalently
$D$-branes intersecting at angles~\cite{Berkooz:1996km, bi} in the
$T$-dual picture.  However, in simple brane constructions the gauge
group sector comes in multiplets of extended supersymmetry, while
matter states are in $N=1$ representations.  In
Ref.~\cite{Antoniadis:2005em} we showed that these economical
string-inspired brane constructions reconcile with unification of
gauge couplings at scales safe from proton decay problems, and provide
a natural Dark Matter candidate.

Indeed a simple way to break supersymmetry in the above context is by
deforming the intersection angles of the Standard Model branes from
their special values corresponding to a supersymmetric configuration.
A small deformation of these angles by $\epsilon$ breaks supersymmetry
via a $D$-term vacuum expectation value (VEV), associated to a
magnetized abelian gauge group factor in the $T$-dual picture,
$\langle D\rangle=\epsilon M_s^2$ with $M_s$ the string
scale~\cite{Bachas:1995ik, Kachru:1999vj}.  This leads to mass shifts
of order $m_0^2\sim \epsilon M_s^2$ for all charged scalar fields
localized at the intersections, such as squarks and sleptons, while
gauginos (and Higgsinos) remain massless. Alternatively supersymmetry
breaking can be communicated to the scalar observable sector by
radiative corrections from a supersymmetric messenger sector, with
$D$-breaking triggered by a magnetized abelian subgroup, or a
non-supersymmetric sector with large extra dimensions.  In all cases
all previously massless scalars in the observable sector are expected
to acquire large masses by radiative corrections and a fine-tuning is
needed in the Higgs sector in order to keep the hierarchy between the
electroweak scale and $m_0$, as required in split supersymmetry.

On the other hand fermion (gaugino and Higgsino) masses are protected
by a chiral $R$-symmetry and the magnitude of radiative corrections
depends on the mechanism of its breaking. In fact $R$-symmetry is in
general broken in the gravitational sector by the gravitino mass but
its value, as well as the mediation of the breaking to the brane
(Standard Model) sector, is model dependent and brings further
uncertainties. Here we will restrict ourselves to possible sources of
fermion mass generation due to brane effects described by open string
propagation within only global supersymmetry, assuming that
gravitational (closed strings) corrections are negligible. Indeed,
$R$-symmetry is in general broken by $\alpha'$-string corrections and
gaugino Majorana masses can be induced by a dimension-seven effective
operator which is the chiral $F$-term~\cite{Antoniadis:2005sd}: $\int
d^2\theta{\cal W}^2{\rm Tr}W^2$, where ${\cal W}$ and $W$ are the
magnetic $U(1)$ and non-abelian gauge superfields, respectively.  Its
moduli dependent coefficient is given by the topological partition
function $F_{(0,3)}$ on a world-sheet with no handles and three
boundaries~\footnote{Two of them correspond to ${\cal W}$ and $W$
gauge groups while the third one can be an orientifold.}.  From the
effective field theory point of view, it corresponds to a two-loop
correction involving massive open string states.  Upon a VEV
$\langle{\cal W}\rangle=\theta\langle D\rangle$, the above $F$-term
generates Majorana gaugino masses that are hierarchically smaller than
the scalar masses and behave as $m_{1/2}^M\sim m_0^4/M_s^3$.

In models where the gauge bosons come in multiplets of extended
supersymmetry, there exists the possibility of generating Dirac
gaugino masses that do not require the breaking of $R$-symmetry. Such
a mass can be induced at one-loop via the effective chiral
dimension-five operator~\cite{Fox:2002bu,Antoniadis:2005em}: $\int
d^2\theta {\mathcal W}{\rm Tr}(WA)$, where $A$ denotes the $N=1$
chiral superfield(s) containing the additional gaugino(s). Upon the
$D$-auxiliary VEV $\langle{\cal W}\rangle$ this term generates Dirac
gaugino masses that scale as $m_{1/2}^D\sim m_0^2/M_s$, and are thus
much higher than the Majorana masses $m_{1/2}^M$. In
Ref.~\cite{Antoniadis:2005em}, we studied the renormalization group
evolution and showed that this scenario is compatible with one-loop
gauge coupling unification at high scale for both cases where the
gauge sector is $N=2$ and $N=4$ supersymmetric.

The low energy sector of these models contains, besides the Standard
Model fields, just some fermion doublets (Higgsinos) and eventually
two singlets, the Binos, if the corresponding corrections to their
Dirac mass vanish. In fact, the Higgsinos must acquire a mass, $\mu$,
of order the electroweak scale in order to provide a Dark Matter
candidate.  This can be induced by the following dimension-seven
operator, generated at one loop
level~\cite{Antoniadis:2005sd,Antoniadis:2005em}: $\int d^2\theta
\mathcal{W}^2 \overline{D}^2{\bar H}_1 {\bar H}_2$, where $H_{1,2}$
are the two $N=1$ Higgs supermultiplets. It follows that the induced
Higgsino mass is of the same order as the gaugino Majorana masses,
$\mu\sim m_{1/2}^M\sim m_0^4/M_s^3$.

The appearance of gauginos in multiplets of extended supersymmetry is
common in previous attempts to embed the Standard Model in
intersecting brane constructions~\cite{Uranga:2005wn}. There are
several examples in the literature of such models with gauge sectors
forming multiplets of either $N=4$~\cite{N=4models} or $N=2$ extended
supersymmetry~\cite{N=2models}.

In this work we describe the general string framework realizing the
above scenario of split supersymmetry with extended supersymmetric
gauge sector and perform an explicit one-loop computation of the
dimension-five and seven effective operators needed to produce Dirac
gaugino and Higgsino masses, respectively. The relevant world-sheet
diagram is the annulus involving two $D$-brane stacks on its
boundaries (or three in the case of Higgsinos).  We find that both
Dirac gaugino and Higgsino masses are in general non-vanishing when
the two brane-stacks are parallel in one of the three internal
compactification planes.  The leading behavior in the supersymmetric
limit, $m_0/M_s\to 0$, gives the coefficient of the corresponding
effective operator, which thus receives non-trivial contributions only
from $N=2$ supersymmetric sectors. Moreover we find that in this limit
the result simplifies and becomes topological; the non-zero mode
determinants cancel and the effective couplings depend only on the
momentum lattice of the plane where the two brane-stacks are parallel.
In the Higgsino case the two fermions should come from an $N=2$
supersymmetry preserving intersection, localized in the remaining two
internal planes.

Finally, for concreteness, we present an explicit string construction
with the Standard Model gauge group, precisely three generations of
quarks and leptons, and sharing the desired features described
above. Moreover, it emerges from an $SU(5)$ grand unified group and
thus satisfies gauge coupling unification, realizing a particular
$D$-brane configuration proposed in
Ref.~\cite{Antoniadis:2004dt}. Standard Model particles live in the
intersection of supersymmetric branes while there is a
non-supersymmetric brane such that particles living in its
intersection with the observable branes are non-chiral and act as
messengers of gauge mediated supersymmetry breaking. In this case the
hierarchy between the string scale and the masses of supersymmetric
partners can be triggered by extra dimensions hierarchically larger
than the string length.

Our paper is organized as follows. In
Section~\ref{section_susy_breaking}, we describe the general framework
of supersymmetry breaking in $D$-brane models intersecting at
angles. In Section~\ref{section_dirac_mass}, we perform the one-loop
computation of the induced Dirac gaugino masses and extract the
coefficient of the relevant dimension-five effective operator by
evaluating the behavior in the supersymmetric limit.  In
Section~\ref{section_higgsino}, we perform a similar computation for
the Higgsino masses and the corresponding dimension-seven effective
operator.  In Section~\ref{model}, we present an explicit construction
of the Standard Model spectrum in this framework with the desired
features.  Our conclusions are drawn in
Section~\ref{section_conclusion} and finally some relevant formulae
are presented in Appendix A and some technical computational details
about the bosonic correlation function of Higgsinos in Appendix B.

\section{\sc Supersymmetry breaking from intersecting branes}
\label{section_susy_breaking}

We will consider a set of intersecting stacks of branes that can be
divided into two subsets: the first one, denoted as ${\cal O}$ (for
observable), gives rise in its light spectrum to the observable
sector, i.e.  a supersymmetric version of the Standard Model with the
gauge sector in $N=2$ or $N=4$ representations. The second subset,
which we denote as ${\cal M}$ (for messenger), provides the
supersymmetry breaking messengers through its intersections with the
branes in ${\cal O}$.

In order to perform explicit computations, we consider a
compactification on a six torus factorizable as $T^2_1 \otimes T^2_2
\otimes T^2_3$ with appropriate projections and orientifold planes to
provide the desired supersymmetric framework. A basis cycles $[{\bf
a}^{(i)}]$ and $[{\bf b}^{(i)}]$ of the corresponding homology classes
is defined for every torus $T^2_i$. Every stack $a$ of $D6$-branes in
type IIA wraps a 3-cycle $[\Pi_a]$ factorizable into the product of
1-cycles:
\begin{equation}
[\Pi_a]=\bigotimes_i [\Pi_a^{(i)}]=\bigotimes_i \left(n_a^{(i)}[{\bf a}^{(i)}]
+m_a^{(i)}[{\bf b}^{(i)}]\right)
\label{jojo}
\end{equation}
and forms angles~\footnote{Here we choose $-\pi/2 \leq
\theta_a^{(i)} \leq \pi/2$.} with the cycles $[{\bf a}^{(i)}]$ given by:
\begin{equation}
\tan {\theta_a^{(i)}}=\frac{m_a^{(i)}R_2^{(i)}/R_1^{(i)}+n_a^{(i)}
\cot\varphi^{(i)}}{n_a^{(i)}}
\end{equation}
where $R_1^{(i)}$ and $R_2^{(i)}$ are the radii along the horizontal
$X^{(i)}$ (${\bf a}^{(i)}$-cycles) and vertical $Y^{(i)}$ (${\bf
b}^{(i)}$-cycles) axes, respectively, and $\varphi^{(i)}$ is the angle
of the tilted torus $T_i^2$.

\begin{figure}[htb]
\centering
\includegraphics[width=0.8\linewidth]{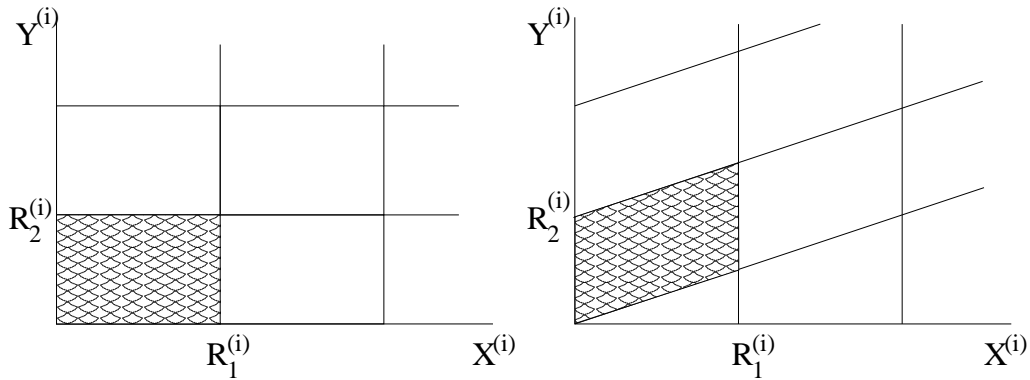}
\caption{\it Rectangular (left panel) and tilted (right panel) tori.}
\label{tori}
\end{figure}

In the presence of an orientifold plane along the $X^{(i)}$-axis, the
angle $\varphi^{(i)}$ is fixed to either $\cot\varphi^{(i)}=0$, which
corresponds to rectangular tori, or to
$\cot\varphi^{(i)}=R_2^{(i)}/2R_1^{(i)}$ which corresponds to tilted
tori (see Fig.~\ref{tori}). 
%
In these cases, we can write the angles
$\theta_a^{(i)}$ as
\begin{equation}
\tan {\theta_a^{(i)}}=\frac{\tilde
m_a^{(i)}}{n_a^{(i)}}\,\frac{R_2^{(i)}}{R_1^{(i)}}
\end{equation}
where we define $\widetilde m_a^{(i)}=m_a^{(i)}+b_i\, n_a^{(i)}$ and
$b_i=0$ ($b_i=1/2$) for rectangular (tilted) tori. From here on and
for simplicity we will remove the tilde from $\widetilde m_a^{(i)}$.
Given a generic couple $(a,b)$ of stacks of branes they intersect with
angles $ \pi \theta_{ab}^{(i)}$ in the $i$-th torus:
\begin{equation}
\theta_{ab}^{(i)} =\frac { \theta_a^{(i)}- \theta_b^{(i)}}{\pi}\
\end{equation}
Of special importance is the number of such intersections as it
measures the number of chiral fermions. It is given by
\begin{equation}
I_{ab} = [\Pi_a]\cdot[\Pi_b] = \prod_i \left(n_a^{(i)}m_b^{(i)} -
m_a^{(i)}n_b^{(i)}\right)
\label{intersection_number}
\end{equation}
and it corresponds in the $T$-dual picture to the index theorem for
compactification with internal magnetic fields. In the special case
where the brane stack $b$ is the image of $a$ under the orientifold
action ($b=a^*$), the chiral states given by this formula fall in two
categories: they transform either in the antisymmetric (A) or in the
symmetric (S) representation of the gauge group, due to the
orientifold projection.  Their respective multiplicities are given by:
\begin{equation}
I_{aa^*}^{A,S}=\left(\prod_i 2m_a^{(i)}\right)
\frac{1} {2}\left(\prod_j n_a^{(j)}\mp 1\right)\, .
\label{intersection_mirrors}
\end{equation}

In simple toroidal compactifications, states arising from open strings
with both ends on the same stack of branes form representations of
$N=4$ supersymmetry. For more complicated compactifications, part of
these states can be projected out and one remains with representations
of lower supersymmetry. In this work, we will focus on the cases where
these states are in $N=4$ or $N=2$ multiplets as already stated.

Open strings stretching between two non-parallel stacks of $N_a$ and
$N_b$ branes give rise to states in the bifundamental representation
$(N_a,{\bar N_b})$ of $U(N_a)\otimes U(N_b)$.  The lightest modes
contain, in addition to massless fermions, scalars with masses:
\begin{eqnarray}
m_{ab,1}^{2}&=& \frac {-|\theta_{ab}^{(1)}|+ |\theta_{ab}^{(2)}|+
  |\theta_{ab}^{(3)}|}{2}M_s^2 \nonumber \\ m_{ab,2}^{2}&=& \frac
{|\theta_{ab}^{(1)}|- |\theta_{ab}^{(2)}|+
  |\theta_{ab}^{(3)}|}{2}M_s^2 \nonumber \\ m_{ab,3}^{2}&=& \frac
{|\theta_{ab}^{(1)}|+ |\theta_{ab}^{(2)}|-
  |\theta_{ab}^{(3)}|}{2}M_s^2 \nonumber \\ m_{ab,4}^{2}&=& \left(1-
\frac {|\theta_{ab}^{(1)}|+ |\theta_{ab}^{(2)}|+
  |\theta_{ab}^{(3)}|}{2}\right) M_s^2
\label{genericmass}
\end{eqnarray}
For parallel branes the first three states become the $N=4$ scalar
partners of the gauge vector bosons. On the other hand, if some of
them are massless there is a boson--fermion degeneracy that indicates
that part of the original supersymmetry is preserved. Moreover, if any
of the $\theta_{ab}^{(i)}$ angles vanishes, a supersymmetric mass
$\ell_i M_s$ can be generated through separation of the branes by a
distance $\ell_i M_s^{-1}$ in the corresponding torus.  Finally, the
last mass in (\ref{genericmass}) contains the contribution of a
massive string oscillator that could become the lightest state for
some values of the angles.

There are three cases corresponding to three different kinds of
intersections that must be discussed:
\vspace{.5cm}

\noindent$\bullet$ The case where branes $a$ and $b$ are in the
observable set $\cal O$ with intersections giving rise to chiral $N=1$
multiplets. The lightest states $\phi_{ab}$ are identified with
quarks, leptons and their supersymmetric partners. This can be
achieved with the choice
\begin{eqnarray}
m_{ab,1}^{2}&=& |\gamma_2|\, M_s^2 \nonumber\\
m_{ab,2}^{2}&=& |\gamma_1|\, M_s^2 \nonumber\\
m_{ab,3}^{2}&=& 0 \nonumber\\
m_{ab,4}^{2}&=& \left(1- |\gamma_3|\right)\, M_s^2 
\label{susymass}
\end{eqnarray}
which is such that for generic values of $\gamma_1$ and $\gamma_2$ the
system only preserves one supersymmetry. For simplicity, we are
using here the notation $\theta_{ab}^{(i)}=\gamma_i$ and choosing
$|\gamma_3|=|\gamma_1|+|\gamma_2|$. 

\vspace{.5cm}
\noindent$\bullet$ The case where branes $a$ and $b$ are in the
observable set $\cal O$, and states in their intersection give rise to
$N=2$ supermultiplets which are identified with supersymmetric pairs
of Higgs doublets. This corresponds to Eq.~(\ref{susymass}) with
$\gamma_1=0$,~i.e.
\begin{eqnarray}
m_{ab,1}^{2}&=& |\gamma_2|\, M_s^2 \nonumber\\
m_{ab,2}^{2}&=& 0\nonumber\\
m_{ab,3}^{2}&=& 0 \nonumber\\
m_{ab,4}^{2}&=& \left(1- |\gamma_2|\right)\, M_s^2 \ .
\label{n2susymass}
\end{eqnarray}
We assume that $n_H$ Higgs chiral multiplets remain light while the
other $N=2$ multiplets get large ($\sim M_s$) supersymmetric masses.
Actually, this is not the only way to obtain Higgs multiplets.  One of
the two doublets may emerge from a chiral intersection together with
the leptons, say between a $U(2)$ and a $U(1)$ brane, while the other
one could arise as a chiral ``anti-doublet" from the intersection of
$U(2)$ with the mirror of $U(1)$ brane.

\vspace{.5cm}
\noindent$\bullet$ The case where brane $a$ is in the observable set
$\cal O$ and brane $b$ in sector $\cal M$. The states $\phi_{ab}$
living at such intersections are assumed to be non-chiral and we will
call them ``messengers'' for reasons that will be apparent below. They
are non-supersymmetric because of mass splitting between fermionic and
bosonic modes. Through loop effects, they will induce supersymmetry
breaking to the observable sector. We will use here the notation
$\theta_{ab}^{(i)}=\alpha_i$ and choose again
$|\alpha_3|=|\alpha_1|+|\alpha_2|$.   Two different kinds of
models can arise at this level. They correspond to the following two
possibilities:

\vspace{.25cm} {\bf i)} The first possibility is that the states in
the intersection $\phi_{ab}$ originate as a perturbation around an
$N=2$ supersymmetric solution.  In fact, in order to have a scale of
supersymmetry breaking hierarchically small compared to the string
scale we perform a tiny deformation of the intersection angles
(preserving $N=2$ supersymmetry):
\begin{equation}
|\alpha_i|\rightarrow |\alpha_i|+\epsilon_i.
\end{equation}
In one of the tori, chosen to be the first one, the branes should
remain parallel and separated~\footnote{Note that this supersymmetric
mass will prevent the appearance of tachyons for scalars in
(\ref{nonsusymass}) and will make the configuration stable.} by a
distance $\ell_1 M_s^{-1}$,~i.e.
\begin{eqnarray}
\alpha_1&=&\epsilon_1=0 \nonumber\\
\alpha_2+\alpha_3&=&\epsilon\, \, \neq 0
\label{alphacond}
\end{eqnarray}
leading to the localized scalar masses:
\begin{eqnarray}
m_{ab,1}^{2}&\simeq &\left( |\alpha_2|\,+ \ell_1^2\right) M_s^2\nonumber\\ 
m_{ab,2}^{2}&\simeq &\left( -\epsilon\,+ \ell_1^2\right) M_s^2\nonumber\\ 
m_{ab,3}^{2}&\simeq & \left( \epsilon\,+ \ell_1^2\right) M_s^2 \nonumber\\
m_{ab,4}^{2}&\simeq &
 \left(1- |\alpha_2|+ \ell_1^2 \right)\, M_s^2
\label{nonsusymass}
\end{eqnarray}
These states will induce at one-loop Dirac masses for gauginos and
Higgsinos, as we will show in next sections. For instance the Dirac
gaugino masses turn out to be
\begin{equation}
m^D_{1/2}\simeq\frac{\alpha}{4\pi}\,\epsilon\,M_s
\label{masagaugino}
\end{equation}
where $\alpha$ is the gauge coupling. This provides a stringy
realization of gauge mediated supersymmetry breaking in the absence of
$R$-symmetry breaking.

Scalars localized at supersymmetric intersections, as those discussed
in (\ref{susymass}), can be given a small supersymmetry breaking
through another tiny deformation of the corresponding angles by
$\epsilon$.  This can be described at the effective field theory level
as a Fayet-Iliopoulos $D$-term breaking corresponding to the presence
of an anomalous $U(1)$ factor $\langle D\rangle\sim\epsilon
M_s^2$~\cite{Bachas:1995ik, Kachru:1999vj}. As a result, scalar masses
in the observable sector can acquire different masses depending on
whether matter is charged under this $U(1)$ or not. In particular:

\begin{itemize}\item
If the observable sector is charged, bosons of the matter multiplets
acquire tree-level masses, $m_0^2\sim \epsilon M_s^2$.

\item
Otherwise the bosons of the observable sector receive masses at the
two-loop level~\cite{Giudice:1998bp}
\begin{equation}
m_0^2\simeq \left(\frac{\alpha}{4\pi}\,{\epsilon}{M_s}\right)^2\ .
\label{masaescalar}
\end{equation}
\end{itemize}

In both cases the transmission of supersymmetry breaking to the
gaugino and Higgsino sector is mediated through the ``messengers''
since the quark and lepton sectors are chiral (and do not contribute
at this order) while the Higgs sector (with masses tuned to remain in
the TeV range) will contribute negligibly.  Note that the brane $b$
does not have to be necessarily in sector $\cal M$.  It could also be
part of the ``observable" sector. Consider for instance the example
where chiral quark doublets come from the intersection of a $U(3)$ and
a $U(2)$ brane stack. Non chiral antiquark doublets, coming from the
intersection of $U(3)$ with the orientifold image of $U(2)$ can play
the role of messengers generating Dirac gaugino masses for both $U(3)$
and $U(2)$ gauge groups. Moreover, the image of $U(2)$ may have non
trivial intersection with another $U(1)$ stack producing leptons, and
thus it is part of sector $\cal O$.

\vspace{.25cm} {\bf ii)} The second possibility corresponds to the
case where a supersymmetry is preserved by the subset of branes in
$\cal O$ for some choice of the compactification moduli, but will
never be conserved by the whole set of branes ${\cal O} \oplus {\cal
M}$.  This is for instance the case for the toroidal compactification
discussed in Section~5 below. Let us denote by $c_i=\pm 1$ the
relevant coefficients that define in (\ref{genericmass}) the
supersymmetry preserved by the observable sector. The intersection
between the ${\cal O}$ and $ {\cal M}$ branes leads to states
$\phi_{ab}$ with a supersymmetry breaking mass
\begin{equation}
m_{ab}^2\simeq \frac{1}{2\pi} \sum_i
c_i\left|\arctan\left\{\frac{m_a^{(i)} R_2^{(i)}}{n_a^{(i)}
R_1^{(i)}}\right\} -\arctan\left\{\frac{m_b^{(i)} R_2^{(i)}}{n_b^{(i)}
R_1^{(i)}}\right\}\right| M_s^2\equiv \epsilon M_s^2
\end{equation}
Keeping the branes $(a,b)$ parallel in one of the tori allows to use
again the states $\phi_{ab}$ at their intersection as messengers to
produce gaugino and Higgsino Dirac masses, as in Sections~3
and~4. However now the desired suppression of $\epsilon$ will be
generated by the presence of a large hierarchy between the size of the
compact dimensions. More precisely for
\begin{equation}
\frac{R_2^{(i)}}{R_1^{(i)}}\ll 1
\end{equation}
one can obtain an $\epsilon$ parameter hierarchically smaller than one
\begin{equation}
\epsilon\simeq \frac{1}{2\pi} \sum_i
c_i\left|\frac{m_a^{(i)}}{n_a^{(i)}}-\frac{m_b^{(i)}}{n_b^{(i)}}\right|\,
\frac{R_2^{(i)}}{R_1^{(i)}}
\end{equation}
where the sum obviously goes over the tori where branes intersect.  In
the absence of tree level supersymmetry breaking, superpartners in the
observable sector will acquire two-loop mass splitting given by
(\ref{masaescalar}). A toy model based on these ideas will be
presented in Section~\ref{model}.

\section{\sc String computation of the Dirac mass} 
\label{section_dirac_mass}

\begin{figure}[htb]
\centering
\epsfxsize=5.5in
\epsfysize=3in
\vspace*{0.2in}
\includegraphics[width=0.8\linewidth]{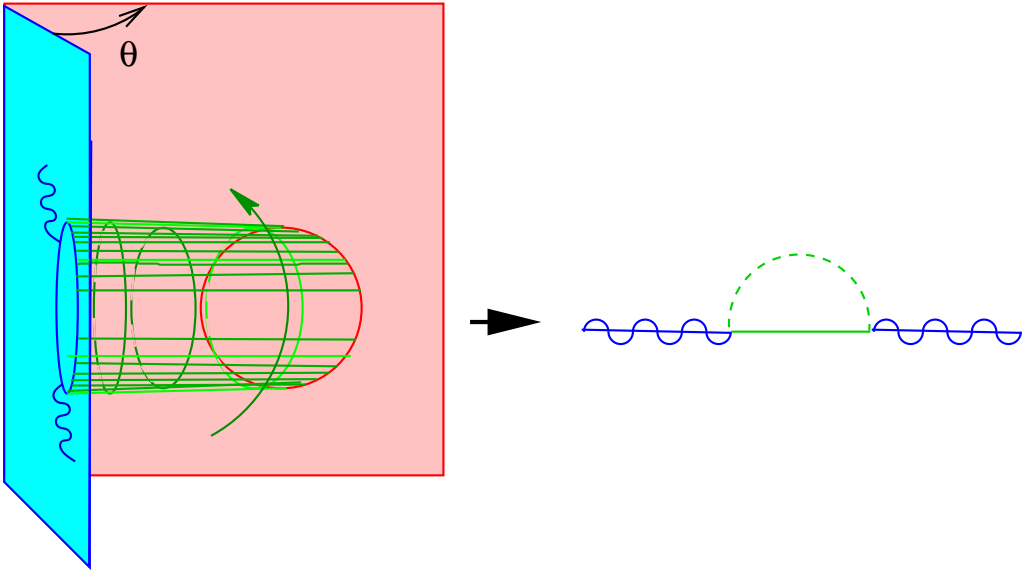}
\caption{\it Non-supersymmetric states stretching between the
two-branes induce at one-loop masses for the gauginos on each brane.}
\label{fig:fig5}
\end{figure}

As the gauginos lie in representations of extended supersymmetry ($N
\geq 2$), they come in copies that can pair up to receive a Dirac mass
at one-loop order in the string genus expansion. The corresponding
world-sheet has the topology of a cylinder stretching between two
stacks of $N_a$ and $N_b$ branes, respectively (see
Fig.~\ref{fig:fig5}).  The vertex operators $V^{(1)}$ and $V^{(2)}$
associated with the gauginos are inserted on one boundary, for example
that corresponding to the $N_a$ branes.

The mass is given by the integrated two-point correlation function of
the relevant vertex operators:
\begin{equation}
A(1,2) = \int dz \int dw \langle V^{(1)}(z) V^{(2)}(w)\rangle \label{amplitude}
\end{equation}
where the integrals are along the boundary of the annulus.
%

The $N=2$ supersymmetry which relates the gauge vector and gauginos is
chosen to be associated with the supercharges that preserve the
(undeformed) $N=2$ preserving brane intersection (see
Eq.~(\ref{nonsusymass}) with $\epsilon=0$). These supercharges are
given by
\begin{eqnarray}
(Q_{\alpha, -++},\ Q^{\dot{\alpha}}_{\ , + - -}) \qquad (Q_{\alpha, ---},
\ Q^{\dot{\alpha}}_{\ , +++} )	\label{conserved_charges}
\end{eqnarray}
where $\pm$ denotes helicities in the internal directions and $\alpha,
\dot{\alpha}$ are Weyl spinor indices in four dimensions. The
supercharges have been grouped into Majorana spinors in four
dimensions. As the gauginos of $N=2$ supersymmetry have the same
internal helicities as the supercharges~(\ref{conserved_charges}), the
vertex operators in the $-\frac{1}{2}$-ghost picture are given by:
\begin{eqnarray}
V^{(1)}_{-\frac{1}{2}}(z) & = & g_o \lambda^1
e^{-\frac{\phi}{2}}e^{\frac{i}{2} \vec S\cdot \vec H}
e^{\frac{i}{2}(H_1 + H_2 + H_3)} e^{i k_1\cdot X} (z) \nonumber \\
V^{(2)}_{-\frac{1}{2}}(z) & = & g_o \lambda^2
e^{-\frac{\phi}{2}}e^{-\frac{i}{2} \vec S \cdot \vec H}
e^{-\frac{i}{2}(-H_1 + H_2 + H_3)} e^{i k_2\cdot X} (z)
\label{gaugino_vertices},
\end{eqnarray}
together with their CPT conjugates which are obtained by simply
reversing helicities in the internal directions and flipping the
space-time chirality. Here $\phi$ is the bosonized two-dimensional
reparametrization ghost field, $X^{\mu}$ are the target space
coordinates, $k_1=k_2=0$ are the space-time momenta, $H_i$ arise from
bosonization of the spin fields and $\vec S$ is the helicity vector in
four dimensions which is constrained by the GSO projection to be
either
\begin{equation}
\vec S = (+,+)  \qquad {\rm or} \qquad \vec S = (-,-)\ .
\end{equation}
Also $g_o$ is some normalization constant and $\lambda^i$ are
Chan-Paton factors of the associated gauge group factor.

On the cylinder, the total $\phi$-charge must vanish. Therefore, we
must use picture-changing to transform one of the vertex operators to
the $+ \frac{1}{2}$ picture. This is done by operating with the BRST
charge:
\begin{equation}
V^{(2)}_{\frac{1}{2}} (w) = \lim_{z \rightarrow w} e^{\phi} T_F(z)
 V^{(2)}_{-\frac{1}{2}}(w)\, ,
\end{equation}
where $T_F(z)$ is the world-sheet supercurrent:
\begin{equation}
T_F(z) = i \sqrt{\frac{2}{\alpha'}} (\partial X^{\mu} \psi_{\mu} +
\sum_{i= 1,2,3} \frac{1}{\sqrt{2}}(\partial Z^i \bar{\Psi}^i + \partial \bar{Z}^i \Psi^i ))\, .
\label{supercurrent}
\end{equation}
The fields $\psi^\mu$ are real world-sheet fermions while in the
second and third terms of~(\ref{supercurrent}) we have used complex
coordinates for the internal directions:
\begin{equation}
Z^i = {X^{(i)} + i Y^{(i)}} \qquad \bar{Z}^i = 
{X^{(i)} - i Y^{(i)}}
\qquad i= 1,2,3
\end{equation}
and analogously for $\Psi^i$ and $\bar{\Psi}^i$. Note that 
$\Psi^i\bar{\Psi}^i=i\partial H_{i}$.

After picture-changing, the vertex operator $V^{(2)}_{-\frac{1}{2}}$
contains several terms, but only one of them contributes to the
two-point function $\langle V^{(1)}(z) V^{(2)}(w)\rangle$. Indeed, the
internal helicities must cancel between the two vertex operators to
yield a non-vanishing correlator. For the choice made
in~(\ref{gaugino_vertices}), $T_F$ must induce a flip of the sign of
$H_1$. We can therefore set

\begin{equation}
V^{(2)}_\frac{1}{2} (w) \rightarrow i \frac{1}{\sqrt{\alpha'}} g_0
\lambda^2 e^{\frac{\phi}{2}} e^{-\frac{i}{2} \vec S \cdot \vec H}
e^{-\frac{i}{2}(H_1 + H_2 + H_3)} e^{i k_2 X} \partial Z^1
\end{equation}

Inserting these operators into a path integral on the cylinder, with
 $k_1 = k_2 = 0$, leads to
\begin{equation}
\langle V^{(1)}(z) V^{(2)}(w)\rangle = i \frac{1}{\sqrt{\alpha'}}
g_o^2 tr(\lambda^1 \lambda^2) N_b \ [BC] \times [FC]
\label{total_correlator}
\end{equation}
where the factor $N_b$ comes from the trace over the Chan-Paton
degrees of freedom of the second boundary of the cylinder. The factors
$[BC]$ and $[FC]$ denote, respectively, correlation functions of
world-sheet bosons and world-sheet fermions together with the
associated ghosts of opposite statistics.  They are given by

\begin{equation}
[BC] = \langle\mathbb{I}\rangle_{bc} \langle\mathbb{I}\rangle_{X^\mu}
 \langle\partial Z^1\rangle_{Z^1}
\langle\mathbb{I}\rangle_{Z^2} \langle\mathbb{I}\rangle_{Z^3}
 \label{BC}
\end{equation}
and
\begin{equation}
[FC] = \sum_{a,b} C\begin{bmatrix} a \\ 
b \end{bmatrix} \langle e^{-\frac{\phi}{2}}
e^{\frac{\phi}{2}}\rangle \langle e^{\frac{i}{2} \vec S \cdot \vec H}
e^{-\frac{i}{2} \vec S \cdot H}\rangle \prod_{i=1,2,3}
\langle e^{\frac{i}{2} H_i} e^{-\frac{i}{2} H_i}\rangle
\label{FC}
\end{equation}
where $\mathbb{I}$ is the identity operator. In the bosonic
correlator, $\langle\cdots\rangle_{Z^i}$ denote bosonic path-integrals
over the complex coordinates $Z^i$ while $\langle\cdots\rangle_{bc}$
is a path integral over the parametrization bc-ghosts. In the
fermionic correlation function the sum is over the spin structures for
which we use the same notation $a,b$ as for brane stacks but the
difference is obvious for the reader.

\subsection{\sc The bosonic correlator}

The one-point functions of identity operators are given by the
appropriate piece of the partition function on the cylinder for the
two stacks of intersecting $D6$-branes.  The bc-ghosts cancel exactly
the path-integral over two of the four space-time coordinates $X^\mu$,
and we are left with the simple
expression:~\footnote{The definitions of the various modular theta
functions appearing in this section are given in the Appendix.}
\begin{equation}
[BC] = \frac{1}{\eta(it)^2} \langle\partial Z^1(w)\rangle_{Z^1}
\frac{\eta(it)}{\vartheta
\begin{bmatrix}
\frac{1}{2} + \alpha_2 \\ \frac{1}{2}\end{bmatrix}}
\frac{\eta(it)}
{\vartheta
\begin{bmatrix}
\frac{1}{2}+ \alpha_3 \\ \frac{1}{2}\end{bmatrix}}
\label{BC_2}
\end{equation}
where $\alpha_2$, $\alpha_3$ are the non-vanishing intersection angles
of the $D6$-branes in the $Z_2$ and $Z_3$ planes.  Here, $t$ is the
open string proper-time parametrizing the world-sheet annulus.

To compute the correlation function of $\partial Z^1(w)$, consider the
coordinate domain of the annulus to be a square parametrized by two
coordinates $(\tau, \sigma)$ with ranges $0 \leq \sigma \leq \pi$, $0
\leq \tau \leq 2 \pi t$ and the identification $\tau \sim \tau + 2 \pi
t $. The operator $Z^1$ is located on the boundary, thus
\begin{equation}
\langle\partial Z^1(w)\rangle = Tr\left(e^{- 2 \pi (t - \tau) H}
\partial Z^1(\tau, 0) e^{-2\pi\tau H}\right) \label{Z3_correlator}
\end{equation} 
where $H$ is the hamiltonian for the $Z^1$ coordinate field.  Only the
zero-modes of the mode expansion of $Z^1$ contribute to the trace
in~(\ref{Z3_correlator}) and, therefore, if the branes would intersect
with a non-trivial angle in the $Z^1$ plane, the correlation function
would vanish. We therefore need to impose
\begin{equation}
\alpha_1 = 0
\end{equation} 
as anticipated in~(\ref{alphacond}). This is also a necessary
condition for the $D$-brane intersection to preserve the
supercharges~(\ref{conserved_charges}).

As the two stacks are parallel in the first torus, they can be
separated at a distance $2 \pi \ell$.  For simplicity, this separation
is taken along the $X^{(1)}$ direction.  The mode expansions read:
\begin{eqnarray}
X^{(1)} & = & x_0^{(1)} + 2 \alpha' p^{(1)} \tau - i \sqrt{2 \alpha'}
\sum_{n \neq 0} \frac{\alpha^{(1)}_{1,n}}{n} e^{i n \tau} \cos{n
\sigma} \nonumber \\ Y^{(1)}& = & y_0^{(1)} + ( 2 n R_2^{(1)} + 2
\ell) \sigma + \sqrt{2 \alpha'} \sum_{n \neq 0}
\frac{\alpha^{(1)}_{2,n}}{n} e^{i n \tau} \sin{n \sigma}
\label{mode_expansions}
\end{eqnarray}
where $n$ is the winding number.  The momentum $p^{(1)}$ is quantized
in units of $1/R_1^{(1)}$.  Using the expression for the Hamiltonian,
\begin{equation}
H = \alpha' (p^{(1)})^2 + \frac{1}{4 \alpha'} (2 n R_2^{(1)} +
2\ell)^2 + \sum_{n > 0, i=1,2} \alpha^{(1)}_{i,-n} \alpha^{(1)}_{i,n}
- \frac{2}{24}
\end{equation}
shows, as stated above, that the oscillators
in~(\ref{mode_expansions}) do not contribute to the trace:
\begin{eqnarray}
\langle\partial Z^1(\tau, 0)\rangle = 2 i \alpha' \sum_{m,n}
\left(\frac{m}{R_1^{(1)}} + \left( \frac{\ell}{\alpha'} + \frac{n
R_2^{(1)}}{\alpha'}\right)\right) e^{- 2 \pi \alpha' t
\left(\frac{m^2}{R_1^{(1)2}} + \left(\frac{n R_2^{(1)}}{\alpha'} +
\frac{\ell}{\alpha'}\right)^2\right)} \frac{1}{\eta(it)^2}
\label{correlator_sum}
\end{eqnarray}
The sum over the quantized momenta does not contribute either, since
the different terms cancel pairwise. However, the sum over windings is
non-vanishing provided that the two stacks are
non-coincident,~i.e.~$\ell \neq 0$.  Inserting~(\ref{correlator_sum})
into~(\ref{BC_2}) yields the complete bosonic correlator:
\begin{equation}
[BC] = \frac{2 i \ap}{\eta^2 \vartheta\begin{bmatrix}\frac{1}{2} + \alpha_2 \\
\frac{1}{2}\end{bmatrix}(0) 
\vartheta\begin{bmatrix}\frac{1}{2} + \alpha_3 \\
\frac{1}{2}\end{bmatrix}(0) }\sum_n \left(\frac{n R_2^{(1)}}{\ap} + 
\frac{\ell}{\ap}\right)
e^{- 2 \pi \ap t \left(\frac{n R_2^{(1)}}{\ap} + \frac{\ell}{\ap}\right)^2}
\label{BC_final}
\end{equation}

\subsection{\sc The fermionic correlator} \label{section_fermion_correlator}

To compute the fermionic correlator, we will make use of the
elementary two-points functions:
\begin{eqnarray}
\langle e^{\frac{\phi}{2}} e^{-{\frac{\phi}{2}}}\rangle 
& = & \left(\frac{\vartheta_1 (z-w)}{\vartheta_1'(0)}\right)^{\frac{1}{4}}\ 
\frac{\eta}{\vartheta\begin{bmatrix} a \\ b \end{bmatrix}(\frac{z-w}{2})}\  
\label{ghost_correlator} \\
\langle e^{\frac{i}{2} H_J} e^{-\frac{i}{2} H_J}\rangle & = &
\left(\frac{\vartheta_1'(0)}{\vartheta_1(z-w)}\right)^{\frac{1}{4}}\
\vartheta\begin{bmatrix} a \\ b
\end{bmatrix}\left(\frac{z-w}{2}\right)\ \frac{1}{\eta} \qquad J=
\xi,1
\label{fermion_correlator1} \\
\langle e^{\frac{i}{2} H_i} e^{-\frac{i}{2} H_i}\rangle & = &
\left(\frac{\vartheta_1'(0)}{\vartheta_1(z-w)}\right)^{\frac{1}{4}}\
\vartheta\begin{bmatrix}a + \alpha_{i} \\ b
\end{bmatrix}\left(\frac{z-w}{2}\right)\ \frac{1}{\eta} \qquad i= 2,3
\label{fermion_correlator2}
\end{eqnarray}
where $\xi$ labels the two non-compact complexified space-time dimensions and $a,b \in
\{ 0, \frac{1}{2} \}$.  The difference
between~(\ref{fermion_correlator1}) and~(\ref{fermion_correlator2})
comes from the fact that in the $Z^2$ and $Z^3$ planes, the brane
intersection angles $\alpha_2$ and $\alpha_3$ are non-vanishing. The
$z-w$ dependence was determined for instance in
Ref.~\cite{Atick:1986ns} and the correct normalization is inferred by
matching the short distance limit $z \rightarrow w$ on both sides.
For example, from the operator product expansion (OPE) of bosonized
fermions we have
\begin{equation}
\langle e^{\frac{i}{2} H_\xi} e^{-\frac{i}{2} H_\xi}\rangle \rightarrow
\frac{1}{(z-w)^{\frac{1}{4}}} \langle\mathbb{I}\rangle =
\frac{1}{(z-w)^{\frac{1}{4}}} \frac{\vartheta\begin{bmatrix} a \\ b
\end{bmatrix}(0)}{\eta}
\label{normalization}
\end{equation}
as the correlator of the identity operator is the spin-structure
dependent partition function of a complex fermion. Taking the same
limit on the right-hand side of~(\ref{fermion_correlator1}) and
comparing with~(\ref{normalization}) allows to check that the
dependence on the cylinder modulus is correctly reproduced.

Inserting~(\ref{ghost_correlator})-(\ref{fermion_correlator2})
into~(\ref{FC}), we obtain:
\begin{equation}
[FC] = \frac{1}{\eta(it)^4} \ \frac{\vartheta'_1(0)}{\vartheta_1(z-w)} \Sigma 
\label{FC_2}
\end{equation} 
where
\begin{equation}
\Sigma = \sum_{a,b} C\begin{bmatrix} a \\ b \end{bmatrix} \vartheta
\begin{bmatrix} a \\ b \end{bmatrix}^2 \left(\frac{z-w}{2}\right)
\vartheta
\begin{bmatrix}
a + \alpha_2 \\
b\end{bmatrix}
\left(\frac{z-w}{2}\right) 
\vartheta\begin{bmatrix}
a + \alpha_3 \\
b\end{bmatrix}
\left(\frac{z-w}{2}\right)\ .
\end{equation}
The coefficients $C\begin{bmatrix} a \\ b \end{bmatrix}$ can be
obtained by the same method as the one used above to deduce the
normalization of the elementary correlators.  From the OPE we deduce
that the fermionic contribution $[FC]$ has a pole in $z-w$ whose
residue, being the correlator of the identity, should be the full
fermionic partition function on the annulus:
\begin{eqnarray}
[FC]& \rightarrow &\frac{1}{(z-w)} Z_{F} = \\
&&\frac{1}{z-w} \sum_{a,b}
\eta_{a,b} e^{- 2 \pi i b (\alpha_2 + \alpha_3)}
\frac{\vartheta\begin{bmatrix} a \\ b \end{bmatrix}^2(0)}{\eta^2}
\frac{\vartheta
\begin{bmatrix}a + \alpha_2 \\ b\end{bmatrix}(0)}{\eta} \frac{\vartheta
\begin{bmatrix}a +\alpha_3 \\ b\end{bmatrix}(0)}{\eta}\nonumber
\end{eqnarray}
where as usual $\eta_{0,0} = \eta_{{\frac{1}{2}},{\frac{1}{2}}} = -
\eta_{{0}, {\frac{1}{2}}}= - \eta_{{\frac{1}{2}},{0}} = 1$.  Comparing
with the limit $z \rightarrow w$ of the right-hand side
of~(\ref{FC_2}), we obtain
\begin{equation}
C\begin{bmatrix} a \\ b \end{bmatrix} = \eta_{a,b} e^{- 2 \pi i b
(\alpha_2 + \alpha_3)}
 \label{coefficients}
\end{equation}

The spin-structure sum in $\Sigma$ can now be performed by using the
Riemann theta-identity~(\ref{riemann}). To apply this identity we must
first use the rearrangement
\begin{equation}
\vartheta\begin{bmatrix}a + \alpha \\ b\end{bmatrix} (z) = e^{2 \pi i
(b + z) \alpha} q^{\frac{\alpha^2}{2} } \vartheta\begin{bmatrix} a \\
b \end{bmatrix}(z + \alpha it)
\label{rearrangement}
\end{equation}
to put the theta functions in the appropriate form. The
$\alpha_i$-dependent phase factor in $C\begin{bmatrix} a \\ b
\end{bmatrix}$ then cancels out. After applying the Riemann identity,
we can use~(\ref{rearrangement}) again to put $\alpha_2$ and
$\alpha_3$ back into the argument of the theta function. The result
is:
\begin{eqnarray}
 \Sigma & =& 2 \vartheta\begin{bmatrix}\frac{1}{2}(1 + \alpha_2 +
\alpha_3) \\ \frac{1}{2}\end{bmatrix} (z - w)\nonumber\\ &&
\vartheta\begin{bmatrix} \frac{1}{2}(1 - \alpha_2 + \alpha_3) \\
\frac{1}{2}\end{bmatrix} (0)\, \vartheta\begin{bmatrix}\frac{1}{2}(1 +
\alpha_2 - \alpha_3) \\ \frac{1}{2}\end{bmatrix} (0)\,
\vartheta\begin{bmatrix}\frac{1}{2}(1 - \alpha_2 - \alpha_3) \\
\frac{1}{2}\end{bmatrix}(0)
\label{sigma}
 \end{eqnarray}
We can now expand this result around the $N = 2$ supersymmetric
configuration of $D$-branes using $\epsilon =\alpha_2 + \alpha_3$.
Expanding~(\ref{sigma}) to lowest order in $\epsilon$ gives
\begin{equation}
[FC] = - 8 \pi^2 \epsilon \eta^2 \vartheta\begin{bmatrix} \frac{1}{2}
+ \alpha_2 \\ \frac{1}{2}\end{bmatrix}(0) \vartheta\begin{bmatrix}
\frac{1}{2} + \alpha_3 \\ \frac{1}{2}\end{bmatrix} (0)
\label{FC_final}
\end{equation}
where we made use of the identity $\vartheta_1'(0)^2 = 4 \pi^2
\eta^6$.

\subsection{\sc The two-gaugino amplitude}

Inserting~(\ref{BC_final}) and (\ref{FC_final})
into~(\ref{total_correlator}) we see that, to lowest order in
$\epsilon$, all contributions from the string oscillators cancel and
only the classical part of the correlation function remains. This
piece is independent of $z-w$ and the integrals over the locations of
the vertex operators in~({\ref{amplitude}}) only contribute a factor
$(2 \pi t)^2$. The result is then simply
\begin{equation}
A(1,2)= 64 \pi^4 t^2 \sqrt{ \ap} \epsilon g_0^2 tr(\lambda^1
\lambda^2) N_b \sum_n \left(\frac{n R_2^{(1)}}{\ap} + \frac{\ell}{\ap}\right)
e^{- 2 \pi t \ap \left(\frac{n R_2^{(1)}}{\ap} + \frac{\ell}{\ap}\right)^2}
\end{equation}
From this we can can obtain the two-gaugino amplitude as
\begin{equation}
\mathcal{A}(1,2) = \frac{1}{4} \frac{V_4}{(8 \pi^2 \ap)^2}
\int_0^{\infty} \frac{dt}{t^3} A(1,2) \label{integrated_amplitude}
\end{equation}
where the integration measure has been obtained by comparing with the
partition function for intersecting $D6$-branes. The infinite volume
factor $V_4$ arises from momentum conservation. Had we chosen $k_1,
k_2 \neq 0$, it would be replaced by a $\delta$-function, $V_4
\rightarrow (2 \pi)^4 \delta^{(4)}(k_1 + k_2)$. The normalization
$g_0$ can in principle be determined. However for our purposes we only
need to know that $g_0$ is proportional to the open string coupling
and therefore $g_0^2 \sim g_s$, where $g_s = e^{\varphi}$ is the
closed string coupling, determined by the VEV of the string dilaton
$\varphi$.  For the Dirac mass of the gauginos we then obtain the
final result:
\begin{equation}
m_{1/2}^D \sim g_s \epsilon N_2 \int_0^{\infty} \frac{dt}{t} \sum_n
\left(\frac{n R_2^{(1)}}{\ap} + \frac{\ell}{\ap}\right) e^{- 2 \pi t
\alpha' \left(\frac{n R_2^{(1)}}{\ap} + \frac{\ell}{\ap}\right)^2}
\label{dirac_mass}
\end{equation}
By a Poisson resummation, it can easily be checked that the integral
is finite. Note also that in the limit where $\ell \rightarrow 0$, the
mass vanishes.

To interpret the meaning of this result, we consider the mass spectrum
of open strings which stretch between the two stacks of $D$-branes. In
the limit where $\epsilon \to 0$, these string modes form $N=2$
hypermultiplets. The mass of the lightest multiplet is given by $\ap
m^2 = \frac{\ell^2}{\ap}$. When $\epsilon$ is non-vanishing,
supersymmetry is completely broken and there is a mass splitting
between fermions and scalars. The lightest fermions remain at $\ap
m_F^2 = \frac{\ell^2}{\ap}$, while their scalar superpartners have
their masses lifted to $\ap m_S ^2 = \ap m_F^2 +
\frac{1}{2}\epsilon$. The result~(\ref{dirac_mass}) then implies the
relation, in the limit $\ell<<\sqrt{\alpha'}$:
\begin{equation}
m^D_{1/2} \sim g_s \epsilon \frac{\ell}{\ap} \sim g_s \frac{m_S^2 -
m_F^2}{M_s^2} m_F\, ,
\end{equation}
where $M_s^2 = 1/\ap$ is the string scale. For string size brane
separation, such that $\ell M_s\sim 1$, we recover the anticipated
behavior (\ref{masagaugino}).

\section{\sc Computation of the Higgsino mass}
\label{section_higgsino}

Consider two stacks of $D6$-branes: the first stack of $N_a$ branes
aligned with the horizontal $X^{(i)}$ axes~\footnote{This can be
achieved by a rotation of the subgroup $U(1)^3\subset SO(6)$.} and the
second stack of $N_c$ branes intersecting the horizontal axes in the
three tori at angles $\theta^{(i)}_c=\beta_i$, $i=1,2,3$.  For $\beta_1=0$
and $\beta_2 = \beta_3$, this configuration preserves $N = 2$
supersymmetry and the lightest string modes on the intersection
describe hypermultiplets. They are made of two $N=1$ chiral
multiplets, identified as two Higgs bosons and their $N=1$
superpartners, the Higgsinos.
\begin{figure}[htb]
\centering
\epsfxsize=5.5in
\epsfysize=3.in
\vspace*{0.2in}
\includegraphics[width=.95\linewidth]{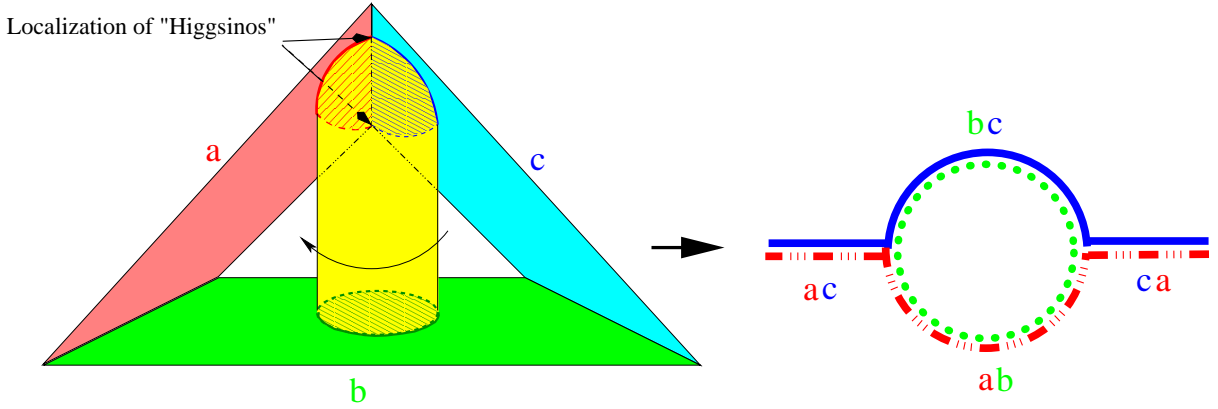}
\caption{\it Higgsinos are localized at the intersection of branes
$a$ and $c$. Non-supersymmetric states stretching between the
branes $(a,b)$ and $(c,b)$ induce at one-loop masses for the Higgsinos.}
\label{fig:fig6}
\end{figure}

A simple way to generate a mass term for these Higgsinos is to
separate the branes in the first torus where they are parallel. This
corresponds to switching on a vacuum expectation value for a
particular scalar that parametrizes the location of the brane in the
direction $X^{(1)}$. The mass is then given by the tension times the
minimal length of the string, which is just the brane separation. This
mass generation is not always possible, as for specific constructions
such as orbifold models, the corresponding scalar is projected out of
the light spectrum.

In this section we will investigate the possibility that the Higgsino
mass is generated at one-loop in a way similar to the Dirac gaugino
mass. The appropriate string amplitude corresponds to a world-sheet
with the topology of an annulus. On the first boundary two vertex
operators that create the Higgsinos are inserted. The Higgsinos arise
from open strings with the two endpoints located on different stacks
of $D$-branes. This means that part of the first boundary of the
annulus lies on the stack of $N_a$ branes and the other part lies on
the other stack of $N_c$ branes (see Fig.~\ref{fig:fig6}). This change
of boundary conditions is implemented by the insertion of twist fields
$\sigma_{\beta_i}$ and $\sigma_{1-\beta_i}$.  The other boundary of
the annulus must be located on a third stack of $N_b$ branes. The two
stacks $N_b$ and $N_a$ intersect at angles $\alpha_i$.  As this
additional stack must preserve the same supersymmetries as the other
stacks, we will start from the simplest possibility $\alpha_i =0$ and
make a small deformation of the angles to break supersymmetry, as
described in Section~\ref{section_susy_breaking}. The generalization
to $\alpha_i\ne 0$ is straightforward.

The vertex operators of the Higgsinos in the $-\frac{1}{2}$-ghost
picture are:
\begin{eqnarray}
V^{(1)}_{-\frac{1}{2}}(z) & = & g_o \lambda^1
e^{-\frac{\phi}{2}}e^{\frac{i}{2} \vec S\cdot \vec H} e^{ \frac{i}{2}
H_1} e^{-i(\beta_2 - \frac{1}{2}) H_2} e^{-i(\beta_3 - \frac{1}{2})
H_3} \sigma_{1-\beta_2} \sigma_{1-\beta_3} e^{i k_1\cdot X} (z)
\nonumber \\ V^{(2)}_{-\frac{1}{2}}(z) & = & g_o \lambda^2
e^{-\frac{\phi}{2}}e^{-\frac{i}{2} \vec S\cdot \vec H} e^{ \frac{i}{2}
H_1} e^{i(\beta_2 - \frac{1}{2}) H_2} e^{i(\beta_3 - \frac{1}{2}) H_3}
\sigma_{\beta_2} \sigma_{\beta_3} e^{i k_2\cdot X} (z) \nonumber
\end{eqnarray}
Note the similarities with the vertex operators of the
gauginos~(\ref{gaugino_vertices}). The helicities have been shifted by
the intersection angles. In addition we have inserted the bosonic
twist operators as world-sheet superpartners of the operators $e^{\pm
i(\beta_i - \frac{1}{2}) H_i}$. In the limit where $\beta_i
\rightarrow 0$, we recover precisely the gaugino vertices since in
this limit the bosonic twists become the identity operators.  Since
the annulus has vanishing ghost-charge we need again to transform one
of the operators to the $\frac{1}{ 2}$-picture. This is done precisely
in the same way as in Section~\ref{section_dirac_mass}. To obtain a
non-vanishing correlator, the world-sheet supercurrent $T_F(z)$ must
be used to flip the sign of $H_1$ in one of the exponents. The result
is
\begin{equation}
V^{(2)}_{\frac{1}{2}}(z) \rightarrow g_0 \lambda^2 e^{\frac{\phi}{2}}
e^{-\frac{i}{2} \vec S \cdot \vec H} e^{-\frac{i}{2}H_1} e^{i(\beta_2 -
\frac{1}{2}) H_2} e^{i(\beta_3 - \frac{1}{2}) H_3} \sigma_{\beta_2}
\sigma_{\beta_3} e^{i k_2 X} \partial Z^1\, .
\end{equation}

Inserting the vertex operators at zero momentum in the correlator
leads to an expression of the same form as~({\ref{total_correlator}),
\begin{equation}
\langle V^{(1)}(z) V^{(2)}(w)\rangle = i \frac{1}{\sqrt{\alpha'}}
g_o^2 tr(\lambda^1 \lambda^2) N_b [BC] \times [FC]
\label{total_correlator2}
\end{equation}
where now
\begin{eqnarray}
[BC] & = & \langle\mathbb{I}\rangle_{bc}
\langle\mathbb{I}\rangle_{X^\mu} \langle\partial Z^1\rangle_{Z^1}
\langle\sigma_{1-\beta_2} \sigma_{\beta_2}\rangle_{Z^2}
\langle\sigma_{1-\beta_3} \sigma_{\beta_3}\rangle_{Z^3} \label{BC_mu}
\\ \nonumber \\ ~ [FC] & = & \sum_{a, b} C\begin{bmatrix} a \\ b
\end{bmatrix} \langle e^{-\frac{\phi}{2}} e^{\frac{\phi}{2}}\rangle
\langle e^{\frac{i}{2} \vec S \cdot \vec H} e^{-\frac{i}{2} \vec S \cdot
\vec H}\rangle \langle e^{\frac{i}{2} H_1} e^{-\frac{i}{2} H_1}\rangle
\nonumber \\ & & \times \prod_{i=2,3}\langle e^{-i (\beta_{i} -
\frac{1}{2}) H_i} e^{i (\beta_{i} - \frac{1}{2}) H_i}\rangle
\label{FC_mu}
\end{eqnarray}
Because of the factor $ \langle\partial Z^1\rangle_{Z^1}$ in $[BC]$
(which was computed in section~\ref{section_dirac_mass}), we must
impose that $\alpha_1=0$ and that the branes are separated by a
distance $\ell \neq 0$ in the $X^{(1)}$ direction in order to obtain a
non-vanishing contribution.

\subsection{\sc The fermion correlator}

The computation of the fermion correlator is similar to the one in
Section~\ref{section_fermion_correlator}.  The correlator is computed
for general $\alpha_2, \alpha_3$ then expanded around the ($N=4$)
supersymmetric configuration $\alpha_i = 0$. A slight generalization
of~(\ref{fermion_correlator2}) is given by:
\begin{equation}
\langle e^{i a H_k} e^{-i a H_k}\rangle =
\left(\frac{\vartheta_1'(0)}{\vartheta_1(z-w)}\right)^{a^2}\ \vartheta
\begin{bmatrix}
a +\alpha_{k} \\ b
\end{bmatrix}
(a (z-w))\ \frac{1}{\eta} \quad ;\quad k= 2,3
\label{fermion_correlator2b}
\end{equation}
with $a,b \in \{ 0, \frac{1}{2} \}$. The remaining elementary
correlators are unchanged and are given in~(\ref{ghost_correlator})
and~(\ref{fermion_correlator1}). Inserting these in~(\ref{FC_mu}) we
obtain
\begin{equation}
[FC] = \frac{1}{\eta^4}
\left(\frac{\vartheta_1'(0)}{\vartheta_1(z-w)}\right)^{1 - \beta_2 (1-
\beta_2) - \beta_3 (1-\beta_3)} \times \Sigma \label{FC_mub}
\end{equation}
where
\begin{equation}
\Sigma = \sum_{a,b} C\begin{bmatrix} a \\ b \end{bmatrix}
\vartheta^2\begin{bmatrix} a \\ b\end{bmatrix}
\left(\frac{z-w}{2}\right) \vartheta
\begin{bmatrix} a + \alpha_2 \\
b\end{bmatrix} \left((\beta_2- \frac{1}{2})(z-w)\right) \vartheta
\begin{bmatrix} a + \alpha_3 \\
b\end{bmatrix}\left((\beta_3 - \frac{1}{2})(z-w)\right)
\end{equation}
The coefficients $C\begin{bmatrix} a \\ b \end{bmatrix}$ are given as
before by~(\ref{coefficients}). Using the Riemann
identity~(\ref{riemann}) allows to perform the spin-structure sum and
put $\Sigma$ into a simpler form.  After imposing the supersymmetry
condition $\beta_2 = \beta_3 \equiv \beta$, one finds:
\begin{eqnarray}
 \Sigma & = & 2 \vartheta
 \begin{bmatrix} \frac{1}{2}(1 + \alpha_2 + \alpha_3)
 \\ \frac{1}{2}\end{bmatrix} (\beta (z - w)) \times \vartheta
 \begin{bmatrix} 
 \frac{1}{2}(1 +\alpha_2 - \alpha_3) \\ \frac{1}{2}\end{bmatrix}(0)
 \nonumber \\ & \times & \vartheta\begin{bmatrix} \frac{1}{2}(1 -
 \alpha_2 + \alpha_3)\\ \frac{1}{2}\end{bmatrix}(0) \times
 \vartheta\begin{bmatrix}\frac{1}{2}(1 - \alpha_2 - \alpha_3) \\
 \frac{1}{2}\end{bmatrix}((1- \beta)(z-w))\, .
\end{eqnarray}
This can now be expanded around the supersymmetric configuration
$\alpha_i = 0$. For simplicity, we take $\alpha_2 = 0$ and $\alpha_3 =
\epsilon$, and expand in powers of $\epsilon$:
\begin{equation}
\Sigma = - 2 \vartheta \begin{bmatrix} \frac{1}{2} \\
\frac{1}{2}\end{bmatrix}(\beta (z-w)) \vartheta \begin{bmatrix}
\frac{1}{2} \\ \frac{1}{2}\end{bmatrix} ((1-\beta)(z-w))
\vartheta_1'(0)^2 \epsilon^2\, .
\label{FC_mu_final}
\end{equation}

\subsection{\sc The bosonic correlator}

The correlator of world-sheet fermions, Eq.~(\ref{FC_mu_final}), is
already of order $\epsilon^2$. Since we are interested only in the
leading $\epsilon$ behavior, for the purpose of computing the bosonic
correlation function $[BC]$, we can keep the leading order
corresponding to $\alpha_i = 0$.  Most pieces of $[BC]$ have already
been computed in Section~\ref{section_dirac_mass}. The additional
complication is the appearance of correlators of bosonic twist
fields. These have been studied for instance for arbitrary genus and
any number of twist fields in Ref.~\cite{Atick:1987kd}. The complete
calculation is presented in Appendix~B where only the relevant results
will be quoted. In fact using Eq.~(\ref{result}) in~(\ref{BC_mu}),
leads to
\bea [BC] & = & i \frac{R_1^{(2)} R_1^{(3)}}{\alpha'} (2 i \alpha')
\left|
\frac{\vartheta_1(z-w)}{\vartheta_1'(0)}\right|^{-2\beta(1-\beta) } 
\frac{1}{\det W(\beta)}
\frac{1}{\eta^8} \nonumber \\ && \times \sum_n \left(\frac{n
R_2^{(1)}}{\alpha'} + \frac{\ell}{\alpha'}\right) e^{- 2 \pi t \alpha'
\left(\frac{n R_2^{(1)}}{\alpha'} + \frac{\ell}{\alpha'}\right)^2}
\nonumber\\ &&\times \sum_{v_1^{(2)}, v_2^{(2)}} e^{-
S_{cl}\left(v_1^{(2)}, v_2^{(2)}\right)} \times \sum_{v_1^{(3)},
v_2^{(3)}} e^{- S_{cl}\left(v_1^{(3)}, v_2^{(3)}\right)}\, ,
 \label{BC_mu_final}
\eea
where $W(\beta)$ is defined as in (\ref{period_matrix}) for the twist
$\beta$ and $v_{1,2}^{(i)}$ are defined in Eq.~(\ref{esta}).

\vskip 0.2cm
\subsection{\sc The Higgsino two-point correlation function}

Finally inserting~(\ref{FC_mub}), (\ref{FC_mu_final})
and~(\ref{BC_mu_final}) into~(\ref{total_correlator2}}) gives for the
two-point function:
\bea \langle V^{(1)}(z) V^{(2)}(w)\rangle &=& 32 i \pi^3 \sqrt{
\alpha'} \epsilon^2 g_0^2 tr(\lambda^1 \lambda^2) N_b \frac{R_1^{(2)}
R_2^{(3)}}{\alpha'} \nonumber\\ &&\sum_n \left(\frac{n
R_2^{(1)}}{\alpha'} + \frac{\ell}{\alpha'}\right) e^{- 2 \pi t \alpha'
\left(\frac{n R_2^{(1)}}{\alpha'} + \frac{\ell}{\alpha'}\right)^2} f(z-w)
\label{higgsino_correlator}
\eea
where
\bea f(z-w) & = & \frac{\vartheta_1(\beta(z-w)) \vartheta_1((1 -
\beta)(z-w)) }{\vartheta_1(z-w) \eta^3}\frac{1}{\det W(\beta)}
\nonumber \\ & &\times \sum_{v_1^{(2)},
v_2^{(2)}} e^{- S_{cl}(v_1^{(2)}, v_2^{(2)})} \times \sum_{v_1^{(3)},
v_2^{(3)}} e^{- S_{cl}(v_1^{(3)}, v_2^{(3)})} \eea

The Higgsino mass $\mu$ is proportional to the two-Higgsino amplitude
obtained upon integration of the correlation
function~(\ref{higgsino_correlator}) over the position of the vertex
operators and inserting the result
into~(\ref{integrated_amplitude}). Since the correlator depends only
on $z-w$, one of the integrals is trivial and contributes simply a
factor $2 \pi t$. The final result reads
\beq \mu \sim g_s \epsilon^2 N_1 \int_0^{\infty} \frac{dt}{t} \sum_n
\left(\frac{n R_2^{(1)}}{\alpha'} + \frac{\ell}{\alpha'}\right) e^{- 2
\pi t \alpha' \left(\frac{n R_2^{(1)}}{\alpha'} +
\frac{\ell}{\alpha'}\right)^2} \ I \eeq
where 
\beq I = \frac{1}{2 \pi t}
\int_0^{2 \pi t} dx f(ix) 
\eeq 
and it is of order $\epsilon^2$ as expected.

\section{\sc A toy model}
\label{model}

In this section we will present a simple model as a framework to
implement the realization of the extended split supersymmetry scenario
described above with gauge bosons in representations of $N=4$
supersymmetry. It consists on a toroidal orientifold based on the
factorized six-dimensional torus $\bigotimes_i T^2_i$ with an
orientifold plane along the $X^{(i)}$ axes. This implies for each
$D6_{a}$-brane with wrapping numbers $(n_a,m_a)$ the presence of the
image brane $D6_{a^*}$ with wrapping numbers $(n_a,-m_a)$.

The minimal setup is given by three intersecting stacks of $D6$-branes
giving rise to the gauge group
\begin{equation}
U(N_{a_1})\otimes U(N_{a_2})\otimes U(N_b) \equiv U(5)\otimes
U(1)\otimes U(N_b)
\end{equation}
The branes $D6_{a_1}$, $D6_{a_2}$ and their images, with
supersymmetric intersections, provide the observable set $\cal O$
while the brane $D6_b$ and its image stand for the (messenger) sector
$\cal M$. The Standard Model gauge group is embedded in $U(5)$ in the
following way:
\begin{itemize}
\item
Open strings localized at the intersection of $D6_{a_1}$ and
$D6_{a_1^*}$ describe three generations transforming as ${\bf 10}$ of
$SU(5)$. Their massless modes transform according to the antisymmetric
representation and come in three generations,~i.e.
\begin{equation}
\prod_i n_{a_1}^{(i)}=1,\quad I_{a_1 a_{1^*}}=3
\end{equation}
It is easy to see that an odd number of generations requires the use
of tilted tori as those described in Fig.~1.
\item
Open strings localized at the intersection of $D6_{a_1}$ and
$D6_{a_2^*}$ provide three generations of ${\bf \overline{5}}$,~i.e.
\begin{equation}
I_{a_1 a_{2^*}}=-3 \ .
\end{equation}
\item
The $D6_{a_2}$ brane being parallel to the $D6_{a_1}$ in one torus,
the massless modes at their intersections will be identified with
$N=2$ hypermultiplets that contain the Supersymmetric Standard Model
Higgs doublets.
\end{itemize}
The supersymmetry breaking messengers arise from strings stretching
between brane $D6_{a_1}$ and $D6_{b^*}$ on one side and between
$D6_{a_2}$ and $D6_{b}$ on the other side. These branes are parallel
(and separated) along the second and first torus, respectively, and
their intersections contain non-chiral matter with non-supersymmetric
masses.

We will discuss here a simple example given in Table~\ref{table2}
where the wrapping numbers of different $D6_I$-branes $(I=a_1,a_2,b)$
in the three tori are listed. It is easy to see that the conditions
for cancellation of RR-charge tadpole are satisfied.
\begin{table}[htb]
\centering
\begin{tabular}{||c|c|c|c||}
\hline
$N_I$ &$(n_I^{(1)},m_I^{(1)})$ & $(n_I^{(2)},m_I^{(2)})$
&$(n_I^{(3)},m_I^{(3)})$ \\ 
\hline\hline 
$N_{a_1}=5$ & $(1,1/2)$ &
$(1,-1/2)$& $(1,-3/2)$ \\ 
$N_{a_2}=1$ & $(1,1/2)$ & $(1,-5/2)$&
$(1,5/2)$ \\ 
$N_b\, n_b^{(1)}n_b^{(2)} n_b^{(3)}=10$
&$n_b^{(1)}(1,1/2)$ & $n_b^{(2)}(1,1/2)$& $n_b^{(3)}(1,1/2)$ \\
\hline
\end{tabular}
\caption{\it Wrapping numbers for the three stacks in the model.}
\label{table2}
\end{table}

The association of any stack of branes $D6_I$ ($I=a_1,a_2,b$) with its
orientifold image $D6_{I^*}$ preserves one of the four
supersymmetries, that we label as $S_\alpha$ with $\alpha=1,
\cdots,4$, if the angles $\theta_I^{(i)}$ satisfy one of the
corresponding relations:
\begin{eqnarray}
S_1: \theta_I^{(1)}+\theta_I^{(2)}+\theta_I^{(3)}=0 && S_2:
-\theta_I^{(1)}+\theta_I^{(2)}+\theta_I^{(3)}=0 \nonumber \\ S_3:
\theta_I^{(1)}-\theta_I^{(2)}+\theta_I^{(3)}=0 && S_4:\, \, \, \, \,
\theta_I^{(1)}+\theta_I^{(2)}-\theta_I^{(3)}=0
\end{eqnarray}
It is easy to check that none of these equations can be simultaneously
satisfied by the three set of branes. Instead, we will look for ratios
of radii
\begin{equation}
A_i=\frac{R_2^{(i)}}{R_1^{(i)}},\quad (i=1,2,3)
\end{equation}
that allow one of the relations, chosen to be $S_1$, to be satisfied
by the stacks $a_1$, $a_2$ and their images. This fixes two of the
ratios as functions of the third one. For instance, in the limit of
small ratios $A_i\ll 1$, the supersymmetric conditions read as
\begin{eqnarray}
A_1&\simeq& \frac{5}{2} A_2\\
A_3&\simeq& \frac{1}{2} A_2
\end{eqnarray}
As a result, the intersections $a_1\cap a_1^*$, $a_2\cap a_2^*$ and
$a_1\cap a_2^*$ preserve $S_1$, the intersections $a_1\cap a_2$ and
$a_1^*\cap a_2^*$ preserve $S_1\oplus S_2$, the intersections $a_1\cap
b$ preserve $S_3\oplus S_4$, and finally the other intersections
involving $b$ and/or $b^*$ do not preserve any supersymmetry. At these
intersections the lightest states have supersymmetry breaking masses
of order:
\begin{equation}
\epsilon M_s^2 \sim A_2 M_s^2
\end{equation}
In particular, the states at the intersection $a_1\cap b^*$ could act
as messengers (a number of $\bf 5+{\bf\bar 5}$ of $SU(5)$) to generate
at one-loop level Dirac masses for fermions, gauginos and Higgsinos,
and at two-loop supersymmetry breaking masses for bosons of the
observable sector.

Note that the gauge symmetry $U(5)$ can be broken to the Standard
Model one by a discrete or continuous Wilson line without introducing
extra fields. Unification of gauge couplings is thus guaranteed in
such particular constructions.  Actually, this model realizes the
particular $D$-brane configuration of Ref.~\cite{Antoniadis:2004dt},
in which the $U(5)$ stack is replaced by two separate brane
collections, $U(3)$ and $U(2)$, describing strong and weak
interactions. Moreover, the hypercharge is the linear combination
$Y=-Q_3/3+Q_2/2$ of the two corresponding $U(1)$ factors $Q_3$ and
$Q_2$. The antisymmetric representation of $SU(5)$ is then decomposed
in terms of the quark doublets, the up antiquarks and the right-handed
lepton, the latter arising as antisymmetric representations of $U(3)$
and $U(2)$, respectively.  Imposing now
$I_{32}=I^A_{33^*}=I^A_{22^*}=3$, the absence of symmetric
representations $\prod_i n^{(i)}=1$ for the two groups, as well as the
absence of antiquark doublets $I_{32^*}=0$, one finds that the strong
and weak branes are parallel in all three planes and thus correspond
to a Wilson line breaking of the GUT group $U(5)$ studied
above~\footnote{Other constructions of a similar model with or without
extended supersymmetric gauge sector are given in
Ref.~\cite{otherconstr}.}.

\section{\sc Conclusions}
\label{section_conclusion}

Small $\epsilon$ deformations of the brane intersection angles provide
a simple mechanism to break supersymmetry, which in the $T$-dual
picture corresponds to the introduction of appropriate combinations of
magnetic fluxes. At the effective field theory level, this can be
described as Fayet-Iliopoulos $D$-term breaking corresponding to the
presence of anomalous $U(1)$ factor(s). Charged scalars localized at
their intersections acquire tree-level masses of order
$m_0^2\sim\epsilon M_s^2$.

Scalars in non-chiral sectors can transmit supersymmetry breaking to
the observable sector through gauge mediated loop
corrections. However, since the $D$-term does not break $R$-symmetry
no gaugino Majorana masses are expected (at least to the lowest
order). Phenomenologically interesting models can appear when the
gauge sector has an extended supersymmetry in which case gauginos can
get Dirac masses. A phenomenological application of this scenario is
given in Ref.~\cite{Antoniadis:2005em} where a minimal model based on
extended split supersymmetry was shown to be consistent with
unification and the presence of Dark Matter.  Dirac masses for
gauginos and Higgsinos where obtained from the $D$-breaking assuming
the presence of some effective higher-dimensional operators. Even if
such operators are absent at tree-level, the present work shows how
they arise at one-loop.  The precise string calculation points out
important features of the messenger sector in the simplest
realization. In particular, the fact that the messengers arise from
strings stretching between branes that are separated in one
direction. They correspond to deformations of $N=2$ sectors.

A simple toroidal compactification, with an orientifold along one
direction, does not allow non-trivial supersymmetric intersecting
branes. A vacuum to be considered as starting point for the above
mentioned $\epsilon$ deformation is then missing, so we instead
propose a different scenario. We identify a subset of supersymmetric
intersecting branes with the observable sector. The other branes lead
to some non-supersymmetric intersections.  The matter living at such
non-chiral intersections will, as in the previous scenario, play the
role of the supersymmetry breaking messenger sector. In order to
obtain masses hierarchically smaller than the string scale, small
ratios of radii are required and they make all intersection angles
small. We have presented a GUT model to illustrate this possibility.

Finally note that for both scenarios there are two possibilities from
the phenomenological point of view: (i) To introduce an $\epsilon$
deformation breaking supersymmetry in the observable sector at
tree-level. (ii) To keep the observable sector supersymmetric at
tree-level.  Non-supersymmetric branes such that the matter living on
their intersection with the observable branes are non-chiral will
transmit supersymmetry breaking by gauge interactions. In case (ii)
both gauginos, squarks, sleptons and Higgses have masses of the same
order of magnitude (as in the usual gauge mediated models) opening up
the possibility of having a supersymmetric spectrum in the TeV range.
This will give rise to a rich phenomenology accessible at LHC
energies.

\section*{\sc Acknowledgments}

Work supported in part by the European Commission under the RTN
contract MRTN-CT-2004-503369, in part by CICYT, Spain, under contracts
FPA2001-1806 and FPA2002-00748, in part by the INTAS contract
03-51-6346 and in part by IN2P3-CICYT under contract Pth 03-1.  We
wish to thank A.~Uranga for useful discussions.  K.B. wishes to thank
the CERN Theory Unit for hospitality. A.D. and M.Q. wish to thank the
LPTHE of ``Universit\'e de Paris VI et VII'' for hospitality. I.A. and
K.B.  acknowledge the hospitality of the ``Universidad Autonoma de
Barcelona'' and IFAE.

\appendix
\section{\sc Appendix A: Theta functions}

In this short appendix, we establish our conventions for the modular
theta functions and list a few useful properties. The theta functions
are defined by:
\begin{equation}
\vartheta\begin{bmatrix} a \\ b \end{bmatrix}(z, \tau) = \sum_{n =
-\infty}^{\infty} e^{\pi i (n+a)^2 \tau + 2 \pi i (n+a)(z+b)}
\end{equation}
where $\tau$ is the (complex) modular parameter of the torus, not to
be confused with the world-sheet coordinate used in the text. On the
cylinder, this parameter is purely imaginary and in the main text we
use the definition $\tau = i t$.

Alternatively, the theta functions can be defined as an infinite
product:
\begin{equation}
\vartheta\begin{bmatrix} a \\ b \end{bmatrix}(z, \tau) = e^{i 2 \pi a
(b+z)} q^{\frac{1}{2}a^2} \prod_{n \geq 1} (1 + q^{n+a - \frac{1}{2}}
e^{2 \pi i (b + z)}) (1 + q^{n - a - \frac{1}{2}} e^{-2 \pi i (b +
z)}) (1 - q^n) \label{theta_defined}
\end{equation} 
where $q = e^{2 \pi i \tau}$.  They satisfy the following periodicity
conditions:
\begin{eqnarray}
\vartheta\begin{bmatrix} a \\ b \end{bmatrix}(z + 1, \tau) & = & -
e^{2 \pi i (a - \frac{1}{2})} \vartheta\begin{bmatrix} a \\ b
\end{bmatrix}(z, \tau) \\ \vartheta
\begin{bmatrix}a \\
b\end{bmatrix} (z + \tau, \tau) & = & 
- e^{- 2 \pi i(b - \frac{1}{2})} e^{- \pi i
\tau - 2 \pi i z} \vartheta\begin{bmatrix} a \\ 
b \end{bmatrix}(z, \tau)\ ,
\label{periodicity}
\end{eqnarray}
which is the reason why they are well suited to describe correlators
on the torus. Defining
\begin{equation}
\vartheta_1(z) \equiv \vartheta
\begin{bmatrix} \frac{1}{2} \\ \frac{1}{2}\end{bmatrix} (z, \tau)
\end{equation}
we can see from the product representation that
\begin{equation}
\vartheta_1(z) = - 2 e^{i \frac{\pi}{4} \tau} sin(\pi z)
\prod_{m=1}^{\infty}(1-q^m)(1-z q^m)(1-z^{-1}q^m).
\end{equation}
In particular
\begin{equation}
\vartheta_1(z) = z \vartheta'_1(0) + \cdots \ , \label{theta1_property}
\end{equation}
a fact that was used repeatedly in the main text.

We also need the Dedekind eta function, which is defined as
\begin{equation}
\eta(\tau) = q^{\frac{1}{24}} \prod_{n \geq 1} (1- q^n)
\end{equation}
It is related to the function $\theta_1(z)$ by the simple identity
\begin{equation}
\theta'_1(0) = - 2 \pi \eta(\tau)^3
\end{equation}

Finally, the theta functions satisfy the following Riemann identity:
\begin{equation}
\sum_{a,b}\eta_{a,b}\vartheta\begin{bmatrix} a \\ b \end{bmatrix}(z_1)
\vartheta\begin{bmatrix} a \\ b \end{bmatrix} (z_2)
\vartheta\begin{bmatrix} a \\ b \end{bmatrix}(z_3) 
\vartheta\begin{bmatrix} a \\ b \end{bmatrix}(z_4)=2
\vartheta\begin{bmatrix} \frac{1}{2} \\  \frac{1}{2} \end{bmatrix}(z'_1)
\vartheta\begin{bmatrix} \frac{1}{2} \\  \frac{1}{2} \end{bmatrix}(z'_2) 
\vartheta\begin{bmatrix} \frac{1}{2} \\  \frac{1}{2} \end{bmatrix}(z'_3) 
\vartheta\begin{bmatrix} \frac{1}{2} \\  \frac{1}{2} \end{bmatrix}(z'_4)
\label{riemann}
\end{equation}
with
\begin{eqnarray}
z'_1= \frac {1}{2}(z_1+z_2+z_3+z_4) && z'_2= \frac
{1}{2}(z_1-z_2+z_3-z_4) \nonumber \\ z'_3 = \frac {1}{2}
(z_1-z_2-z_3+z_4) && z'_4= \frac {1}{2}(z_1+z_2-z_3-z_4)\nonumber
\end{eqnarray}


\section{\sc Appendix B: Bosonic correlator for Higgsino mass}

Let us consider the correlator on the torus with twists of an
arbitrary angle $\theta$. This correlator has a quantum piece
$\mathcal{Z}_{qu}$ and a classical piece which takes into account the
contributions from world-sheet instantons:
\begin{equation}
\langle\sigma_{1-\theta}(z_1) \sigma_{\theta}(z_2) \rangle =
\mathcal{Z}_{qu} \sum_{i} e^{-S_{cl}(i)}\, ,
\label{twist1}
\end{equation}
where the sum is over classical solutions of the equations of
motion. Both pieces are expressed in terms of the so-called
cut-differentials, which are holomorphic one-forms on a branched
covering of the punctured torus. When expressed as functions of the
torus coordinate $z$, they become multi-valued with branch-cut
singularities at the location of the punctures. In the present
context, where the torus has only two punctures corresponding to the
insertions of the two twist fields, these cut-differentials are simply
given by:
\begin{eqnarray}
\omega(z) = \vartheta_1(z - z_1)^{-\theta}
\vartheta_1(z-z_2)^{-(1-\theta)} \vartheta_1(z - z_2 - \theta(z_1 - z_2))
\\ \nonumber \\ \omega'(z) = \vartheta_1(z - z_1)^{-(1-\theta)}
\vartheta_1(z-z_2)^{-\theta} \vartheta_1(z - z_1 + \theta(z_1 - z_2))
\label{cut_differentials}
\end{eqnarray}
and satisfy
\begin{equation}
\omega(z + 1) = \omega(z + \tau) = \omega(z) \qquad \omega'(z+1) =
\omega'(z+\tau) = \omega'(z)
\end{equation}
where $\tau$ is the complex modulus of the torus. The singularities at
the insertions $z_i$ depend on the twist-angle $\theta$.

From these cut-differentials, we construct a ``period matrix" $W_a^i$
defined as
\begin{equation}
W^1_a = \int_{\gamma_a} dz\ \omega(z) \qquad W^2_a = \int_{\gamma_a}
d\bar{z}\ \bar{\omega}'(\bar{z}) 
\label{period_matrix}
\end{equation}
where the paths $\gamma_a $, $a=1,2$, denote the canonical homology
basis of the torus. From~\cite{Atick:1987kd} we obtain the quantum
part of the correlation function:
\begin{equation}
\mathcal{Z}_{qu} = \frac{1}{\det W} |\vartheta_1(z_1 -
z_2)|^{-2\theta(1-\theta)}
\end{equation}
To determine the classical part of the correlator, we need to evaluate
the action for all possible classical solutions of the equations of
motion. Each contribution will have the form
\begin{equation}
S_{cl} = \frac{1}{4 \pi \ap} \int d^2z\ (\partial Z_{cl}
\bar{\partial}\bar{Z}_{cl} + \bar{\partial}Z_{cl} \partial
\bar{Z}_{cl}) \label{classical_action}
\end{equation}
where $Z_{cl}$ and $\bar{Z}_{cl}$ are classical
solutions,~i.e.~$Z_{cl}$ solves $\partial \bar{\partial} Z_{cl}=0$ on
the punctured torus with appropriate boundary conditions (i.e.~branch
cut singularities) at the location of the punctures. In particular,
under parallel transport along the canonical homology basis of the
torus, $Z_{cl}$ is expected to shift as
\begin{equation}
\Delta_a Z_{cl} = \int_{\gamma_a} dz \partial Z_{cl} + 
\int_{\gamma_a} d \bar{z} \bar{\partial}Z_{cl} = v_a\, ,
\end{equation}
where $v_a$ are given complex numbers. For instance for a toroidal
compactification $Z= X^1+ i X^2$ with
\begin{equation}
X^1 \sim X^1 + 2 \pi R_1 \qquad X^2 \sim X^2 + 2 \pi R_2
\end{equation}
when $Z_{cl}$ is transported along a closed loop on the world-sheet
torus, it must return to its original value modulo a lattice vector of
the space-time torus. This implies
\begin{equation}
v_1 = 2 \pi (m_1 R_1 + i n_1 R_2) \qquad v_2 = 2 \pi (m_2 R_1 + i n_2
R_2)\, ,
\label{vi_torus}
\end{equation}
where the integers $m_i$ and $n_i$ correspond to the winding numbers
of the closed string propagating in the loop.

The classical part of the correlator is written as
\begin{equation}
\mathcal{Z}_{cl} = \sum_{v_1, v_2} e^{-S_{cl}(v_1, v_2)}
\end{equation}
where $S_{cl}(v_1, v_2)$ is now expressed in terms of the period
matrix~(\ref{period_matrix}):
\begin{equation}
S_{cl} = \frac{i}{4 \pi \ap} v_a \bar{v}_b \left( (\bar{W}^{-1})^b_{\
1} \bar{W}^1_{\ d} M^{ad} + (W^{-1})^a_{\ 2} W^2_{\ c} M^{bc}\right)
\label{classical_actionb}
\end{equation}
The two-point correlation function of twist fields on the torus is
then given by
\begin{equation}
\langle\sigma_{1-\theta}(z_1) \sigma_{\theta}(z_2)\rangle_T=
 \frac{1}{\det W} \left|\vartheta_1(z_1 -
 z_2)\right|^{-2\theta(1-\theta)} \sum_{v_1, v_2} e^{-S_{cl}(v_1,
 v_2)} \label{twist_correlator}
\end{equation}
where the sum over the $v_i$ corresponds to a sum over all possible
winding modes.

The case we are interested here, the annulus two-point correlation
function, is obtained from~ (\ref{twist_correlator}) by taking the
``square root" and replacing the complex modulus $\tau$ by $\tau
\rightarrow it$. As a result, the twist correlator on the annulus
takes the form
\begin{equation}
\langle\sigma_{1-\theta}(z_1) \sigma_{\theta}(z_2)\rangle_A = K(\tau)
\frac{1}{\det^{\frac{1 }{ 2}} W} \left| \frac{\vartheta_1(z_1 -
z_2)}{\vartheta_1'(0)}\right|^{-\theta(1-\theta)} \sum_{v_1, v_2}
e^{-S_{cl}(v_1, v_2)}
\label{twist_correlator_annulus}
\end{equation} 
where now
\begin{equation}
S_{cl} = \frac{i}{8 \pi \ap} v_a \bar{v}_b \left( (\bar{W}^{-1})^b_{\ 1}
\bar{W}^1_{\ d} M^{ad} + (W^{-1})^a_{\ 2} W^2_{\ c} M^{bc}\right)
\label{classical_actionc}
\end{equation}
and $K(\tau)$ is a normalization factor which can depend on the
modulus. It is now understood that the twist fields are located on one
of the boundaries of the cylinder, $z_j = i y_j$ with $y_j$ real.  The
values of $v_a$ must also be appropriately modified. Instead
of~(\ref{vi_torus}), we now have
\beq
v_1 = 2 \pi R_1 m \qquad v_2 = i 4 \pi (R_2 n + \ell)
\label{esta}
\eeq
where $R_1, R_2$ are the compactification radii of the space-time
torus and $\ell$ is the separation between the $D$-branes along the
$X^2$ direction.

Taking the limit $y_1 \rightarrow y_2$, allows to check that this
expression is correct and to determine the normalization constant
$K(\tau)$. In this limit, the operator product expansion of the twist
fields behaves as:
\begin{eqnarray}
\langle\sigma_{1-\theta}(z_1) \sigma_{\theta}(z_2)\rangle &\to &
\frac{1}{(z_1 - z_2)^{ \theta(1-\theta)}}
\langle\mathbb{I}\rangle\\ & =& \frac{1}{(z_1 - z_2)^{
\theta(1-\theta)}} \frac{1}{\eta^2} \sum_{m,n} \exp\left(- 2 \pi \ap t
\left(\frac{m^2}{R_1^2} + \left(\frac{nR_2}{\ap} +
\frac{\ell}{\ap}\right)^2\right)\right)\nonumber
\label{limit_lhs}
 \end{eqnarray}
where we have used that $\langle\mathbb{I}\rangle$ is the partition
function on the annulus of the complex boson $Z$. The factor
$1/\eta^2$ is the contribution of the oscillators while the sum is
over the momentum and winding modes of the string. The same result
should be obtained by taking the limit $z_1 \rightarrow z_2$ on the
right-hand side of~(\ref{twist_correlator_annulus}).

The expression~(\ref{cut_differentials}) for the cut-differential
implies that
\begin{equation}
\omega(z) \rightarrow 1 \qquad \omega'(z) \rightarrow 1
\end{equation}
In this limit the components of the period matrix and its inverse can
be explicitly evaluated :
\begin{equation}
 W = \begin{pmatrix} W^1_{\ 1} & W^1_{\ 2} \\ W^2_{\ 1} & W^2_{\ 2}
 \end{pmatrix} = \begin{pmatrix} it & 1 \\ -it & 1 \end{pmatrix}
 \qquad W^{-1} = \frac{1}{2i t} \begin{pmatrix} 1 & -1 \\ it & it
 \end{pmatrix} \label{explicit_period_matrix}
\end{equation}
Inserting this in the action~(\ref{classical_actionc}) leads to
\begin{equation}
S_{cl} = - \frac{\pi}{2 \ap t} R_1^2 m^2 - \frac{2 \pi t}{\ap}(R_2 n + \ell)^2
\end{equation}
so that the classical piece of the correlator becomes
\begin{eqnarray}
\sum_{v_1, v_2} e^{-S_{cl}} & = & \sum_m \exp\left(- \frac{\pi
R_1^2}{2 \ap t} m^2\right) \ \sum_n \exp\left(-2 \pi t \ap
\left(\frac{\ell}{\ap} + n \frac{R_2}{\ap}\right)^2\right) \nonumber
\\ & = & \left(\frac{2 \ap t}{R_1^2}\right)^{\frac {1}{2}} \sum_{m,n}
\exp\left(- 2 \pi \ap t \left(\frac{m^2}{R_1^2} +
\left(\frac{nR_2}{\ap} + \frac{\ell}{\ap}\right)^2\right)\right)
\label{explicit_classical_part}
\end{eqnarray}
where the second equality is obtained by a Poisson resummation.

Using~(\ref{theta1_property}) and~(\ref{explicit_period_matrix}), the
limit $z_1 \rightarrow z_2$ of the quantum part is
\begin{equation}
\mathcal{Z}_{qu} = \frac{K(\tau)}{|z_1 - z_2|^{\theta (1- \theta)}}
\left(\frac{1}{2it}\right)^{1/2}\, .
 \label{explicit_quantum_part}
\end{equation}
Inserting~(\ref{explicit_classical_part})
and~(\ref{explicit_quantum_part}) into~(\ref{twist1}) and comparing
with~(\ref{limit_lhs}) shows that both expressions agree to each other
provided that $K(\tau) = \left(i R_1^2/\ap\right)^{1/2} \, 1/\eta^2$.

The final result for the correlation function of twist-antitwist on
the annulus is
\begin{equation}
\langle\sigma_{1-\theta}(z) \sigma_\theta(w)\rangle = \left(\frac{i
R_1^2}{\ap}\right)^{\frac{1}{ 2}} \frac{1}{\eta(it)^2}
\frac{1}{\det^{\frac{1}{ 2}} W}
\left|\frac{\vartheta_1(z-w)}{\vartheta_1'(0)}\right|^{-\theta(1-\theta)}
\sum_{v_1, v_2} e^{-S_{cl}(v_1, v_2)}
\label{result}
\end{equation}
where $v_1$ and $v_2$ are given in Eq.~(\ref{esta}).


\end{document}